\documentclass[twocolumn,amssymb, aps, prb,dvipdfmx]{revtex4-2}

\usepackage[dvipdfmx]{graphicx}
\usepackage{amsmath}
\usepackage{braket}
\usepackage{bm}
\usepackage{amssymb,amsthm}
\usepackage{float}
\usepackage{color}
\usepackage{tikz}

\begin{document}

\title{Angular-time evolution for the Affleck-Kennedy-Lieb-Tasaki chain and its edge-state dynamics}%

\author{Koutaro Nakajima}%
\email[]{f21j005d@mail.cc.niigata-u.ac.jp}
\affiliation{Graduate School of Science and Technology, Niigata University, Niigata 950-2181, Japan}
\author{Kouichi Okunishi}%
\affiliation{Department of Physics, Niigata University, Niigata 950-2181, Japan}
\date{\today}%

\begin{abstract}
We study the angular-time evolution that is a parameter-time evolution defined by the entanglement Hamiltonian for the bipartitioned ground state of the Affleck-Kennedy-Lieb-Tasaki (AKLT) chain with the open boundary.
In particular, we analytically calculate angular-time spin correlation functions $\langle S_n^\alpha(\tau)S_n^\alpha(0)\rangle$ with $\alpha = x,y,z$, using the matrix-product-state (MPS) representation of the valence-bond-solid state with edges.
We also investigate how the angular-time evolution operator can be represented in the physical spin space with the use of gauge transformation for the MPS.
We then discuss the physical interpretation of the angular-time evolution in the AKLT chain.
\end{abstract}
\maketitle
%\tableofcontents

\section{Introduction}
Quantum entanglement in quantum many-body systems has attracted much attention since it often plays a key role in understanding the physics behind exotic behaviors induced by strong quantum fluctuations.
For instance, entanglement entropy logarithmically diverges at the quantum critical point, which is characterized by the central charge of a conformal field theory \cite{Vidal2003}.
Also, degeneracy structures of the entanglement spectrum\cite{Li2008} have been often used for detecting the symmetry-protected topological (SPT) order.
 A typical example is the $\mathbb{Z}_2 \times \mathbb{Z}_2$ symmetry breaking in the valence-bond-solid (VBS) state for the Affleck-Kennedy-Lieb-Tasaki (AKLT) chain\cite{AKLT,AKLT2,KT_cmp1992,KT_prb1992}, which also played an essential role in resolving the Haldane conjecture of the $S=1$ Heisenberg chain\cite{Haldane1983} and its entanglement entropy was exactly calculated for various setups.\cite{Korepin2004, Korepin2010,Geraedts2010}
In general, nevertheless, the entanglement entropy and entanglement spectrum are not conventional observables, and thus it is a difficult task to directly measure them in a realistic experimental situation of quantum many-body systems.

An interesting hint on how to extract information of quantum entanglement from quantum many-body systems is involved in the Unruh effect for the vacuum state of relativistic quantum field theories, where the quantum entanglement between the left and right Rindler wedges is in principle detectable as a thermalized spectrum by a constantly accelerating observer through the auto-correlation function of a quantum field with respect to the proper time.\cite{Fulling1973, Unruh1976, RMP_Unruh}
In analogy with the Unruh effect, it was recently shown that the angular-time spin correlation functions can also be used as a protocol of detecting the entanglement spectrum for the bipartite partitioning of the XXZ chain.\cite{Okunishi2019} 
A key point of the lattice version of the Unruh effect is that the entanglement Hamiltonian of the XXZ chain corresponds to a lattice Lorentz boost operator based on the integrability,\cite{Sogo1983, Thacker1986} implying that the angular-time evolution defined by the entanglement Hamiltonian basically plays a similar role to the proper-time evolution in the original Unruh effect for the quantum field theory.

However, the lattice Lorentz boost does not imply a naive acceleration of an observer for the spin system.
Moreover, the analytic relation between the usual $S^{z}$-diagonal bases for the physical spins and the entanglement-Hamiltonian diagonalizing bases, which is a quantum-spin-system counterpart of the Bogoliubov transformation for the quantum field theory, is not easy to understand even for the XXZ chain.
Thus, it remains unclear how we can observe the real-time dynamics of angular-time evolved spins in a quantum spin chain under a realistic experimental situation.
Our motivation is to clarify the physical interpretation of the angular-time evolution by generalizing it to a typical quantum spin system where the relationship between the angular-time evolution and the usual real-time one can be tractable, which may be also relevant to the entanglement physics of quantum many-body systems.

\begin{figure}
\includegraphics[width=6cm]{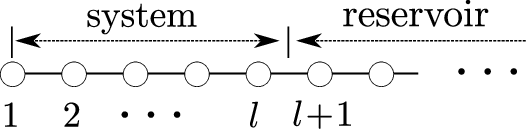}
\caption{Setup of the system and reservoir parts in a half-infinite chain.
The left edge corresponds to the $i=1$ site, and the system part contains $l$ sites from the left edge. }
\label{fig_setup}
\end{figure}

In this paper, we address the angular-time evolution problem for the AKLT chain.
The AKLT chain has been providing fundamental understandings of quantum entanglement in the SPT-ordered state \cite{Pollmann2010, XieGuWen2011}, and its edge states are expected to be a possible resource of measurement-based quantum computation\cite{Miyake2008,Miyake2010}.
Then, we consider a half-infinite VBS state with open boundary conditions and then explicitly construct the eigenstates for the reduced density matrix by tracing out the bulk reservoir part (See Fig. \ref{fig_setup}),  using the matrix-product-state (MPS) representation\cite{Fannes1989, Fannes1992, Klumper1993, Schollwock2011}.
On the basis of the exact transformation between the physical $S=1$ spins and the entanglement-Hamiltonian diagonalizing bases, we define the angular-time spin correlation functions, which exhibit an oscillating behavior with the frequency specified by the entanglement spectrum, and further discuss the physical interpretation of the angular-time evolution in the AKLT chain. 
We then find that the angular-time evolution for the VBS state can be generated by a weak magnetic field applied in the $S=1$ spins in the system part, which would provide essential insight relevant to experimental studies of the quantum entanglement in quantum many-body systems.

This paper is organized as follows.
In the next section,  we briefly review the MPS representation of the VBS state and explain our setup of the reduced density matrix.
In particular, we explicitly construct the basis set that diagonalizes the reduced density matrix using the MPS, which is essential in analyzing the angular-time evolution problem. 
In Sec. 3, we define the angular time based on the entanglement Hamiltonian and then calculate angular-time spin correlation functions. 
In Sec. 4, we extract the angular-time evolution operator in the physical $S=1$ spin space with the help of the gauge transformation in the MPS and then discuss its physical interpretation.
Finally, we summarize our results and discuss their relevance to an ESR experiment for the $S=1$ Heisenberg chain.

\section{Valence-bond-solid state and reduced density matrix}
\label{Sec2}

In this section, we briefly review the setup of the MPS description of the AKLT chain and the VBS state\cite{AKLT,AKLT2} required for discussion about the angular-time evolution.
The Hamiltonian of the AKLT chain is defined as
\begin{align}
    \mathcal{H}_\mathrm{AKLT}&=\sum_{i=1}^{N-1} \bm{S}_{i}\cdot\bm{S}_{i+1}+\frac{1}{3}(\bm{S}_{i}\cdot\bm{S}_{i+1})^{2} \, ,
\label{eq_aklt}
\end{align}
where $\bm{S}_i$ is $S=1$ spin operators at $i$ site.
Moreover, the open boundary condition at $i=1$ and $N$ is assumed for Eq. (\ref{eq_aklt}), though the half-infinite chain limit($N\to \infty$) is basically considered later.

In Eq. (\ref{eq_aklt}), a significant point is that the two-body interaction works as the projection operator onto the total spin $J=2$ of the $i$ and $i+1$ sites, so that its ground state is exactly constructed as a VBS state.
Using the MPS representation, then, we can compactly write down the VBS state with fixed edges as
\begin{align}
    \ket{\Psi_\mathrm{VBS}} &= c_N  \sum_{\{\sigma\}}|\sigma_1 \cdots \sigma_N  \rangle   \mathrm{v}^{\dagger} {A}^{\sigma_1}  {A}^{\sigma_2} \cdots   { A}^{\sigma_N} \Gamma \mathrm{v}' \, ,
\label{eq_VBS}
\end{align}
where $\sigma_i  \in 0$, $\pm1$ denotes the index for a $S=1$ spin state at $i$ site and $c_N$ is a normalization constant.
The matrix $A^{\sigma}$ is explicitly written as
\begin{align}
   &A^{1}=\begin{bmatrix}
        0&\sqrt{\frac{2}{3}}\\
        0&0
    \end{bmatrix},\,
    A^{0}=\begin{bmatrix}
        \frac{-1}{\sqrt{{3}}}&0\\
        0&\frac{1}{\sqrt{3}}
    \end{bmatrix},\,
    A^{-1}=\begin{bmatrix}
        0&0\\
        -\sqrt{\frac{2}{3}}&0
    \end{bmatrix} \, ,
\end{align}
and edge vectors in the auxiliary space are also introduced as 
\begin{align}
    &\mathrm{v}=
    \begin{bmatrix}
        \cos\theta \\
        \sin\theta
    \end{bmatrix}, \quad
    \mathrm{v}'=
    \begin{bmatrix}
        \cos\theta^{\prime} \\
         \sin\theta^{\prime}
    \end{bmatrix}\, ,
\end{align}
where $\theta$ and $\theta'$ denotes real angle variables specifying the edge states at $i=1$ and $N$ respectively.
Note that $\sum_\sigma {A^\sigma}^\dagger A^\sigma = 1$.
In addition,   
\begin{align}
     \Gamma \equiv \begin{bmatrix}
        0&-1\\
        1&0
    \end{bmatrix}
\end{align}
which originates from the dangling singlet pair of the auxiliary spins at the right edge.
Then, $|\Psi_\mathrm{VBS} \rangle$ with appropriate $\theta$ and $\theta'$ provides the four degenerating ground states of Eq. (\ref{eq_aklt}) with the open boundaries. 

Let us introduce the diagrammatic representation of the MPS, which is very useful to visualize complicated contraction of tensor legs.
 For instance, we illustrate elements of the local tensor $A$ as
\begin{align}
A_{\mu\mu'}^\sigma = \, \vcenter{ \hbox{\includegraphics[scale=0.35]{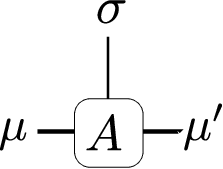}} }
\: ,
\end{align}
where the horizontal tick line indicates the leg indices $\mu$ and $\mu'$ for the auxiliary spin degrees of freedom.
Then, the expansion coefficient of the VBS state in Eq. (\ref{eq_VBS}) can be diagrammatically represented as
\begin{align}
\langle \sigma_1 & \cdots \sigma_N|\Psi_\mathrm{VBS} \rangle =\notag \\
 & c_N \times
\vcenter{ \hbox{\includegraphics[scale=0.35]{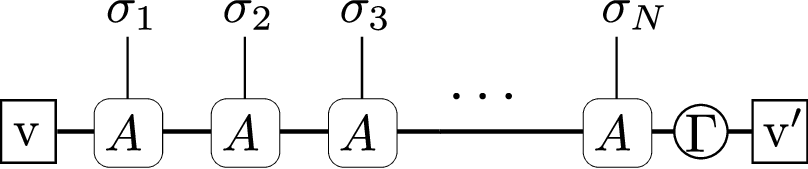}} }
\label{eq_diavbs} 
\end{align}
where the connected line represents the contracted indices with respect to auxiliary spins and the box and circle symbols respectively denote the edge vectors and the $\Gamma$ matrix.   

As in Fig. \ref{fig_setup}, we regard the region of length $l$ from the left edge as a system part and its complement as the reservoir part.
After contracting the spins in the reservoir part for Eq. (\ref{eq_VBS}), we take the half-infinite chain limit, $N \to \infty$.
Then, we obtain the reduced density matrix for the system part as
\begin{align}
  &\bra{\sigma_{1}\cdots\sigma_{l}} \rho^{(l)}\ket{\sigma^{\prime}_{1}\cdots\sigma^{\prime}_{l} }  \nonumber \\
  &  =  \mathrm {v}^{\dagger}  {A^{\sigma_{1}}} \cdots A^{\sigma_{l}}  {A^{\sigma^{\prime}_{l}}}^\dagger  \cdots { A^{\sigma^{\prime}_{1}} }^\dagger   \mathrm{v}.
  \label{eq_rdm_s1}
\end{align}
Although the matrix size of $\rho^{(l)}$ is $3^l \times 3^l$, its matrix rank is always reduced to 2 corresponding to the auxiliary spin degrees of freedom. 

In order to diagonalize $\rho^{(l)}$, we explicitly set up the bases with the use of the MPS for the system part: 
\begin{align}
    | q \rangle\!\rangle_{}&=\sum_{\sigma} \ket{\sigma_{1}\cdots\sigma_{l}} 
    \; 
\vcenter{ \hbox{\includegraphics[scale=0.35]{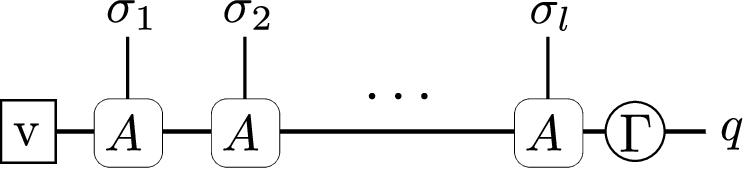}} } \quad ,
    \label{eq_qbase}
\end{align}
where $q \in \uparrow, \downarrow $ corresponds to the index of the auxiliary spin emerging at the leg of $\Gamma$ at the right edge.
Hereafter, we assign the double-braket symbol $| \cdots \rangle\!\rangle$ for the ket state in the 2-dimensional auxiliary spin space.

As well known for the VBS state, $|\! \uparrow\rangle\!\rangle$ and $|\! \downarrow \rangle\!\rangle_{}$ may not be orthonormal; The overlaps for them are explicitly obtained as  
\begin{align}
    &\xi_{\uparrow\uparrow} \equiv {}_{}\langle\!\langle \uparrow\!|\! \uparrow \rangle\!\rangle_{}=\frac{1}{2}-\frac{\gamma^{l}}{2}\cos2\theta  \, ,\\
    &\xi_{\downarrow\downarrow}\equiv  {}_{} \langle\!\langle \downarrow\!|\!\downarrow \rangle\! \rangle_{}= \frac{1}{2}+\frac{\gamma^{l}}{2}\cos2\theta\, ,\\
    &\xi_{\uparrow\downarrow} \equiv {}_{}\langle \!\langle \uparrow\!|\!\downarrow \rangle \!\rangle_{} = -\frac{\gamma^{l}}{2}\sin2\theta \, ,
\end{align}
where $    \gamma \equiv -\frac{1}{3} $. 
Here, we note $\xi_{\downarrow\uparrow}= \xi_{\uparrow\downarrow}$ for the real wavefunction.
Then, the Gram-Schmidt orthonormalization for $| \uparrow, \downarrow \rangle\!\rangle_{}$ yields 
\begin{align}
    |v_1 \rangle \!\rangle = \frac{|\!\uparrow \rangle\!\rangle_{} }{ \sqrt{\xi_{\uparrow\uparrow}} }  \, , \quad 
    |v_2 \rangle \!\rangle = \frac {   |\downarrow \rangle\!\rangle_{}  
    -  \xi_{\uparrow\downarrow} / \xi_{\uparrow\uparrow} |\!\uparrow \rangle\!\rangle_{} }{  \sqrt{ \xi_{\downarrow\downarrow} -  \xi_{\uparrow\downarrow}^2 / \xi_{\uparrow\uparrow}}  } \, ,
    \label{eq_vbase}
\end{align}
Using the orthonormal bases of Eq. (\ref{eq_vbase}), we have the reduced density matrix represented in the auxiliary space as 
\begin{align}
    \tilde{\rho}^{(l)}&=
    \begin{bmatrix}
        \xi_{\uparrow\uparrow}+ \frac{\xi_{\uparrow\downarrow}^2}{\xi_{\uparrow\uparrow}} &   \frac{\xi_{\downarrow\downarrow}\xi_{\uparrow\downarrow}}{\sqrt{\xi_{\uparrow\uparrow}\xi_{\downarrow\downarrow} - \xi_{\uparrow\downarrow}^2 } }\\
         \frac{\xi_{\downarrow\downarrow}\xi_{\uparrow\downarrow}}{\sqrt{\xi_{\uparrow\uparrow}\xi_{\downarrow\downarrow} - \xi_{\uparrow\downarrow}^2 } } & \xi_{\downarrow\downarrow}-\frac{\xi_{\uparrow\downarrow}^2}{\xi_{\uparrow\uparrow}} 
    \end{bmatrix} \, ,
\end{align}
and then obtain the eigenvalues
\begin{align}
    \lambda_{\pm}=\frac{1\pm \gamma^{l}}{2} \, ,
\end{align}
which is consistent with the well-established result for the AKLT chain.\cite{Korepin2010,Schollwock2011}
Note that $\lambda_\pm$ do not contain $\theta$, implying that the bipartition entanglement of the half-infinite AKLT chain with the open boundary is independent of the left edge.
Also,  we notice that  the states belonging to  $3^l -2$ number of zero eigenvalues involved in Eq. (\ref{eq_rdm_s1}) defined in the $S=1$ spin space has been projected out by Eq. (\ref{eq_qbase}).

In order to discuss the angular-time evolution, an important object is the eigenvectors of $\rho^{(l)}$, which explicitly provide the transformation between the physical $S=1$ spin space and the $\rho^{(l)}$-diagonalizing bases in the auxiliary space.
We write the eigenstates corresponding to $\lambda_\pm$ as $| \nu \rangle \!\rangle$ with $\nu=\pm$, respectively.
Then, we can formally write 
\begin{align}
    |\nu  \rangle  \!\rangle = \sum_q |q  \rangle\!\rangle \, G_{q\nu} \, .
    \label{eq_gtrans}
\end{align}
where $q = \uparrow, \downarrow$ for Eq. (\ref{eq_qbase}) and the matrix elements of $G$ are explicitly obtained as 
\begin{align}
    G_{\uparrow + }&= - \sqrt{\frac{2}{1+\gamma^l}} \sin \theta \, , \quad    
    G_{\uparrow -} = \sqrt{\frac{2}{1-\gamma^l}}\cos\theta \\
    G_{\downarrow +}&= \sqrt{\frac{2}{1+\gamma^l}}\cos \theta \, ,\quad
    G_{\downarrow -}= \sqrt{\frac{2}{1-\gamma^l}}\sin\theta \, .
\end{align}
Here, note that $G$ may not be an orthogonal matrix if $|q\rangle\!\rangle_{}$ is not orthonormal bases. 
Combining Eqs. (\ref{eq_qbase}) and (\ref{eq_gtrans}), we can diagrammatically represent $|\nu\rangle\!\rangle$ in the MPS form as
\begin{align}
| \nu \rangle\!\rangle = %\notag \\
% & 
\; 
 & \sum_{\{\sigma\}}|\sigma_1 \cdots \sigma_l \rangle \; 
\vcenter{ \hbox{\includegraphics[scale=0.35]{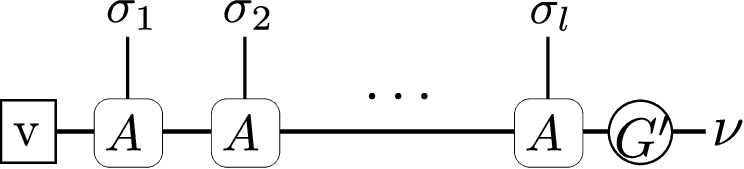}} } \; ,
\label{eq_bogoliubov}
\end{align}
where we have introduced $G'\equiv \Gamma G$ for later convenience.

\section{Angular-time Correlations}

On the basis of the setup in Sec. \ref{Sec2}, we discuss the angular-time evolution for the AKLT chain.
The term ``angular time" was originally introduced to the angular variable of the Rindler coordinate for 1+1 dimensional quantum field theories, which corresponds to the proper time of an accelerating observer in the Unruh effect.
An essential point is that an infinitesimal shift along the ``angular time" direction is generated by the Lorentz boost operator ${\cal K}$ \cite{Rindler1966, Fulling1973, Brazinikov1998}, with which the reduced density matrix for the half-infinite bipartitioning is also written as  $\rho \propto \exp(- {\rm const}\, {\cal K} )$.
In other words,  ${\cal K}$ plays both the roles of the generator of the angular time evolution and the entanglement Hamiltonian.
Then,  the Unruh-DeWitt detector captures the entanglement spectrum for the bipartitioned vacuum state through the angular-time correlation function, where the Bogoliubov transformation diagonalizing the angular-time evolution operator or equivalently the entanglement Hamiltonian relates particles observed by the accelerating observer to those in the usual Minkowski space-time.\cite{Unruh1976,DeWitt, Birrell,Brout,RMP_Unruh}

For the XXZ chain, recently, a quantum-spin-system version of the angular-time evolution was introduced as a parameter time evolution based on the lattice Lorentz boost operator\cite{Sogo1983, Thacker1986} equivalent to the entanglement Hamiltonian, where the angular-time spin correlations were investigated numerically with the use of the density matrix renormalization group.\cite{Okunishi2019}
However, the relation between the angular-time evolution of spins and its real-time spin dynamics has not been clarified yet, because the quantum-spin-chain counterpart of the Bogoliubov transformation in the original Unruh effect is not obtained analytically even for the XXZ chain.
Here, it should be noted that the entanglement Hamiltonian for quantum spin systems is also explored recently in the context of the discretized version of the Bisognano-Wichmann theorem.\cite{Dalmonte2018,Guidici2018, BW1975}
%This fact has prevented us from developing a deeper understanding of the lattice-Unruh effect.

Now we discuss a generalization of the angular-time evolution for the AKLT chain.
An essential observation is the above-mentioned role of the entanglement Hamiltonian;
Although the lattice Lorentz boost operator is not available in general, we can directly define the angular-time evolution
\begin{align}
 {\cal U}(\tau)= e^{-i\tau {\cal K} }\, ,
\label{eq_atime}
\end{align}
in terms of the entanglement Hamiltonian that is well defined as
\begin{align}
     \mathcal{K} \equiv -\log\rho\, ,
     \label{eq_aHam}
\end{align}
where $\rho$ is the reduced density matrix for a bipartitioned wavefunction.
Here, note that the angular-time $\tau$ in Eq. (\ref{eq_atime}) is introduced as a parameter time through ${\cal K}$, in contrast to the original Unruh effect for the quantum field theory where $\tau$ corresponds to the proper time for an accelerating observer.
Then, we find out that the spin-chain counterpart of the Bogoliubov transformation is available as Eq. (\ref{eq_bogoliubov}).
In the following, we first investigate angular-time spin correlations, which are a quantum-spin-chain version of the Wightman function associated with the Unruh-Dewitt detector for a relativistic quantum field.

\subsection{angular-time spin correlations}

Let us consider the angular-time evolution for the AKLT chain.
Suppose that $l$ is the length of the system part and $n$ is the index of a certain site in it.
The reduced density matrix $\rho^{(l)}$ of Eq. (\ref{eq_rdm_s1}) as well as ${\cal K}^{(l)}$ are defined on the $3^l$ dimensional Hilbert space of the $S=1$ spins in the system part.
We then write the expectation value of a certain operator ${\cal O}$ in the Hilbert space of $S=1$ spins as
\begin{align}
 \langle {\cal O} \rangle \equiv \mathrm{Tr}[ \rho^{(l)} {\cal O} ] \, .
\end{align}
In particular, the angular-time spin correlation, which plays a central role in this paper, is  defined as 
\begin{align}
   & \langle S^{\alpha}_{n}(\tau)S^{\alpha}_{n}(0)\rangle  =   \mathrm{Tr} [ \rho^{(l)}S^{\alpha}_{n}(\tau)S^{\alpha}_{n}(0) ]
\label{eq_atcorr}
\end{align}
where 
\begin{align}
    S^{\alpha}_{n}(\tau)& \equiv e^{i \tau \mathcal{K}^{(l)} }S^{\alpha}_{n}e^{-i\tau \mathcal{K}^{(l)} } \, ,
\label{eq_spin1evol}
\end{align}
represents the angular-time evolution of $S=1$ spin operators with $\alpha \in x,y,z$ for $ 1 \le n \le l$ in the Heisenberg representation. 
We note that Eq. (\ref{eq_atcorr}) corresponds to the quantum-spin-chain counterpart of the Wightman function in the original Unruh effect for the quantum field theory.

In calculating Eq. (\ref{eq_atcorr}),  we need its spectral decomposition.
Then, an important point is that  ${\cal K}^{(l)}$ is eventually defined with the spectrum of $\tilde{\rho}^{(l)}$,
\begin{align}
\omega_\pm = - \log \lambda_\pm \, .
\end{align}
This implies that ${\cal K}^{(l)}$ contains $3^l -2$ number of infinitely oscillating modes corresponding to  the zero eigenvalues of ${\rho}^{(l)}$, reflecting that the $|q\rangle\!\rangle$ basis is served as a projection to the 2-dimensional auxiliary space.
Although  the complete set in the S=1 spin space is required for the exact calculation, the $3^l -2$ number of highly oscillating modes would not affect the physical part of the angular-time spin correlation.
In other words, we may assume an approximate relation
\begin{align}
  \sum_{\{ \sigma \}  } | \sigma_1 \cdots \sigma_l \rangle  \langle \sigma_1 \cdots \sigma_l | 
   \simeq \sum_{\nu = +, -} |\nu \rangle\!\rangle \langle\!\langle \nu|\, ,  
   \label{eq_approx_trans}
\end{align}
for a practical calculation of Eq. (\ref{eq_atcorr}) through Eqs. (\ref{eq_qbase}) and (\ref{eq_gtrans}).
Also, Eq. (\ref{eq_approx_trans}) works as a projection operator for the system part onto the 2-dimensional auxiliary space.
Then, the diagrammatic representation of the angular-time spin correlations is illustrated as
\begin{align}
    \langle S^{\alpha}_{n}(\tau) & S^{\alpha}_{n}(0)\rangle  = \notag \\  
&\; 
\vcenter{ \hbox{\includegraphics[scale=0.35]{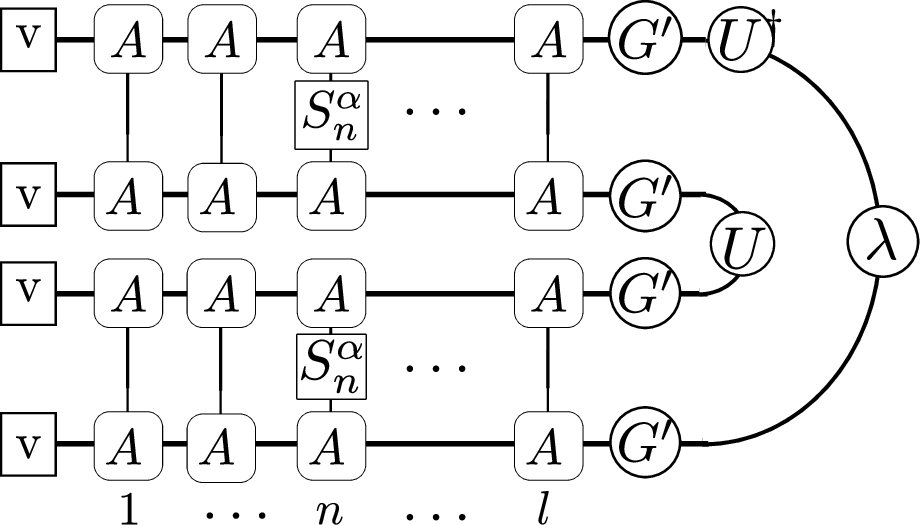}} } \; ,
\label{eq_diagram_atc}
\end{align}
where 
\begin{align}
U \equiv 
\begin{bmatrix} 
e^{-i\tau \omega_+} & 0 \\ 0 & e^{-i\tau \omega_-}
\end{bmatrix}\, ,
\end{align}
denotes the angular-time evolution in the $|\nu\rangle\!\rangle$ basis.

\begin{figure}[bt]
    \centering
     \includegraphics[width=7.8cm]{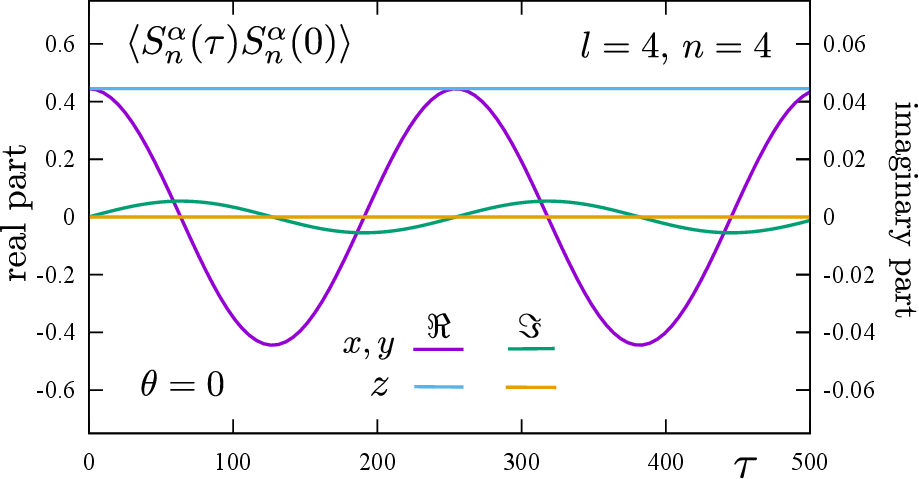} 
      \caption{Angular-time spin correlations for $l=4$ and $n=4$ at $\theta=0$.
      The real and imaginary parts of the angular-time spin correlations for $\alpha=x$ and $y$ exhibit oscillating behavior. The frequency of the oscillation is $\Omega_4 \simeq 0.0247$ (The period is $T\simeq 254$). While those for $\alpha=z$ have no $\tau$ dependence. 
     We note that the amplitude of the oscillation $D_{l,n}$ decays rapidly, as $n$ becomes away from the entangle point ($n=l$). 
      }
      \label{fig_corr_0}
\end{figure}

Using matrix elements $\langle \!\langle \nu| S_n^\alpha |\nu' \rangle\!\rangle $ in Appendix \ref{app_A},  
we calculate Eq. (\ref{eq_diagram_atc}). 
The result is summarized as follows.
\begin{align}
    \langle S^{x}_{n}(\tau)S^{x}_{n}(0)\rangle  & = C_{l,n} \sin ^2(2 \theta ) \nonumber \\ 
     + &D_{l,n} \cos^{2}(2\theta) (\cos\Omega_l\tau+i\gamma^{l}\sin\Omega_l\tau) \, , \label{eq_aciix}\\
    \langle S^{y}_{n}(\tau)S^{y}_{n}(0)\rangle
    &=D_{l,n}(\cos\Omega_l\tau+i\gamma^{l}\sin\Omega_l\tau)\, , \label{eq_aciiy}\\
    \langle S^{z}_{n}(\tau)S^{z}_{n}(0)\rangle  & =  C_{l,n}\cos ^2(2 \theta )\nonumber\\
    + &D_{l,n} \sin^2(2\theta) (\cos\Omega_l\tau+i\gamma^{l}\sin\Omega_l \tau) \label{eq_aciiz} \, ,
\end{align}
where  
\begin{align}
\Omega_l = \omega_+ -\omega_- = \log \frac{1+\gamma^l}{1-\gamma^l} \, ,
\label{eq_Omega}
\end{align}
and 
\begin{align}
C_{l,n}&=\frac{2 \left(\gamma^{n}-\gamma^{l-n +1}\right)^2}{1+ \gamma^l} + \frac{2\left(\gamma^{ n}+\gamma^{l-n+1}\right)^2}{1- \gamma^l}\, ,
\label{eq_C_ln} \\
D_{l,n}& = \frac{4\gamma^{ 2(l-n+1)}}{1-\gamma^{2l}} \, .
\label{eq_D_ln}
\end{align}
Figure \ref{fig_corr_0} shows a typical example of the angular-time spin correlations for $\theta=0$ and $l=4$, where oscillating behavior of $\langle S_n^{x}(\tau) S_n^x(0) \rangle$ and $\langle S_n^{y}(\tau) S_n^y(0) \rangle$ is clearly observed.
Here, we should note that although  Eqs. (\ref{eq_aciix}) and (\ref{eq_aciiz}) have $\theta$ dependence,  $\langle S_n^{x}(\tau) S_n^x(0) \rangle + \langle S_n^{z}(\tau) S_n^z(0) \rangle = C_{l,n} + D_{l,n}(\cos\Omega_l\tau+i\gamma^{l}\sin\Omega_l\tau)$ is independent of $\theta$.
This reflects the rotation around the $y$ axis does not disturb the $\mathbb{Z}_2 \times \mathbb{Z}_2$ symmetry in the VBS state.
Thus we can discuss general features of the angular-time evolution in the result for $\theta=0$.

The above results demonstrate that $\tau$ dependence of the angular-time spin correlations exhibits oscillating behavior with the frequency $\Omega_l$,  from which one can extract information about the entanglement spectrum embedded in the VBS state.
An important feature of $\Omega_l$ is that it depends only on the length $l$ of the system part. 
In the VBS state, the entanglement originates from the SPT order reflecting the $\mathbb{Z}_2 \times \mathbb{Z}_2$  symmetry, which gives rise to a characteristic time scale of $\Omega_l$ specified only by $l$.
Then,  Eq. (\ref{eq_Omega}) clarifies that $\Omega_l$ captured by the angular-time spin correlations arises from the splitting of the spectrum induced by the edge effect. 
Actually, we have a moderately large $\Omega_l$ for a relatively small $l$. 
As $l$ increases, however, the entanglement spectrum exhibits double degeneracy, and thus $\Omega_l \sim 2\gamma^{l}$.
This implies that ${\cal U}(\tau)$ is reduced to be an overall phase factor in the bulk region ($l\gg 1$).
Note that this behavior is consistent with the fact that the edge-spin distribution in the Haldane state with the open boundaries rapidly decays toward the bulk region\cite{Kennedy1990,Miyashita1993,White1993}. 
Thus, the edge-induced splitting of the entanglement spectrum is essential for the angular-time evolution in the AKLT chain, which is distinct from the XXZ chain that has site-dependent frequencies reflecting the infinite series of the spectrum.\cite{Okunishi2019}

We next discuss the amplitude of the angular-time spin correlations. 
The $n$ dependence of the oscillation amplitude $D_{l,n}$ is read off from  Eq. (\ref{eq_D_ln}), which decays exponentially as $n$ becomes distant from the entangle point ($n=l$). 
Of course, the angular-time spin correlation near the entangle point captures the strong entanglement between the system and reservoir parts.
For the AKLT chain, the correlation length is given by $\xi = -1/\log |\gamma| = 0.91\cdots$, and the decay rate of the amplitude is also governed by this value. 
Thus, a dominant contribution to observing the angular-time spin correlations arises from a few sites from the entangle point.

Here, we comment on a way of evaluating the validity of the approximated relation (\ref{eq_approx_trans}) in the angular-time correlations.
As mentioned before, the highly oscillating modes corresponding to the zero eigenvalues of $\rho^{(l)}$ could be canceled with each other in the angular-time spin correlations.
At $\tau=0$, however, such cancellation never occurs and thus $ \bm{S}_n^2 = 2$ should be hold in principle.    
On the other hand, we obtain the angular-time correlation function in the $\tau\to 0$ limit as  
\begin{align}
 \sum_\alpha \langle (S^{\alpha}_{n}(0) )^2\rangle = C_{l,n} + 2 D_{l,n} \,,
 \label{eq_corr_tau0}
\end{align}
which is less than the exact magnitude of $ \bm{S}_n^2  = 2$.\cite{example_value}
In other words, the deviation of Eq. (\ref{eq_corr_tau0}) from $ \bm{S}_n^2 = 2$ provides a quantitative check of Eq. (\ref{eq_approx_trans}) in the angular-time spin correlation.
For the case of $l=n=4$ in Fig. \ref{fig_corr_0}, about $30$\% 
of the spin magnitude is lost in Eq. (\ref{eq_corr_tau0}).
As the site index $n$ becomes away from the entangle point,  Eq. (\ref{eq_corr_tau0}) also decays exponentially. 
This suggests that the physical role of Eq. (\ref{eq_approx_trans}) is to project out the spin fluctuations irrelevant to the quantum entanglement across the entangling point.
However, note that Eq. (\ref{eq_approx_trans}) does not affect $\Omega_l$.

\subsection{angular-time correlations for magnetization}

\begin{figure}[bt]
    \centering
      \includegraphics[width=7.8cm,keepaspectratio]{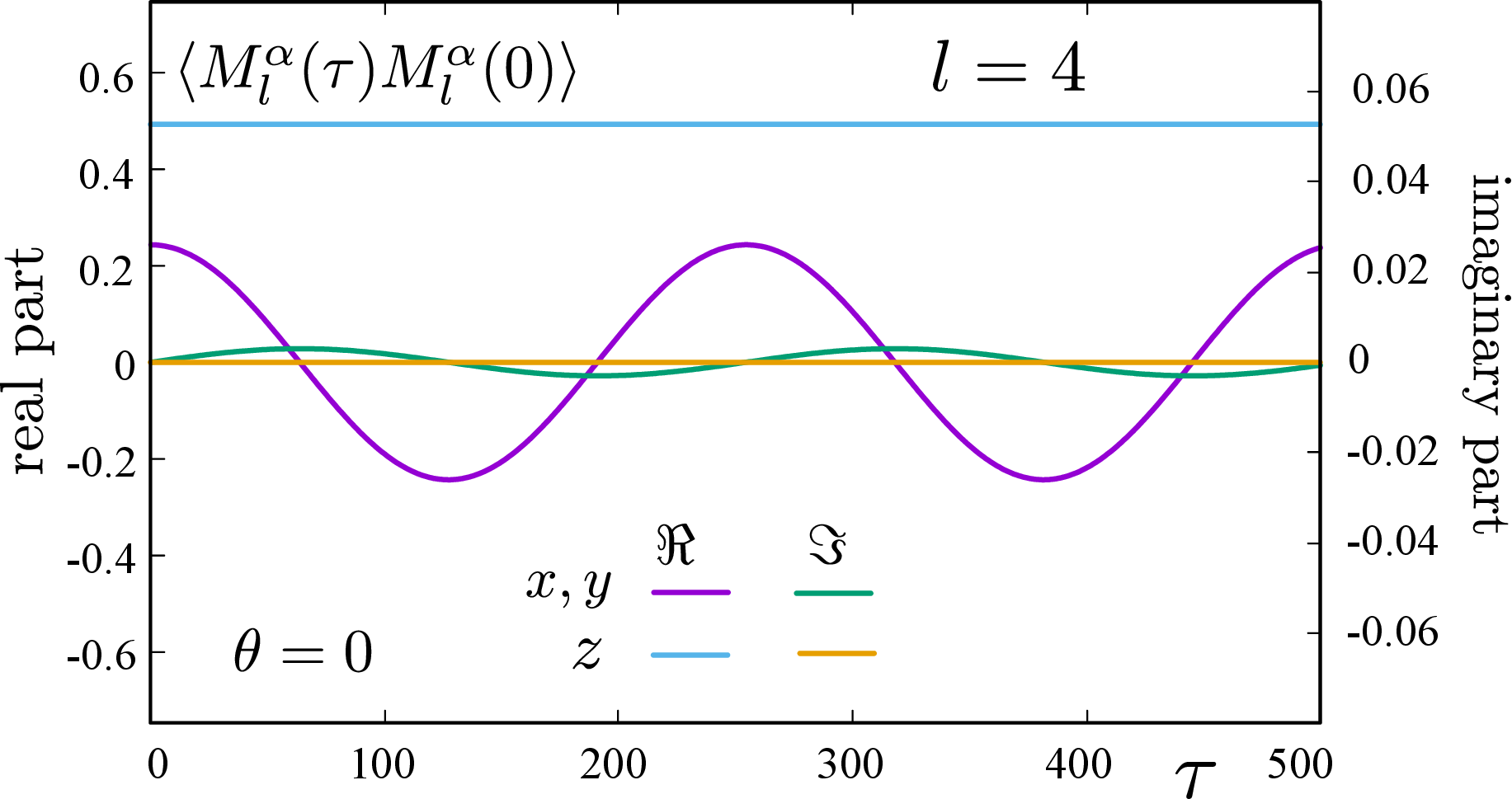}
      \caption{
      Angular-time correlations of $M^\alpha_l(\tau)$ for $l=4$ at $\theta=0$.
      The real and imaginary parts of the angular-time spin correlations for $\alpha=x$ and $y$ exhibit oscillating behavior. The frequency of the oscillation is identical to that in Fig. \ref{fig_corr_0}. Meanwhile, the amplitude of the oscillation rapidly approaches $ 1/4$ as $l$ increases.}
\label{fig_ac_mag}
\end{figure}

So far, we have focused on the $n$-dependent angular-time spin correlations, which are a quantum-spin-chain counterpart of the Wightman function for a quantum field observed by a constantly accelerating observer in the original Unruh effect. 
However, site-selective measurement of spins is usually difficult in a realistic condensed matter experiment.
We next consider the total magnetizations in the system part, 
\begin{align}
M_l^\alpha  \equiv \sum_{n=1}^l S^\alpha_n \, ,
\end{align}
and calculate their angular-time correlations $\langle M_l^\alpha(\tau) M_l^\alpha(0)\rangle = \sum_{m,n}^{l}\langle S^{\alpha}_{m}(\tau)S^{\alpha}_{n}(0)\rangle$.
Using the result of $\langle S^{\alpha}_{m}(\tau)S^{\alpha}_{n}(0)\rangle$ presented in Appendix \ref{sec_appdx2}, we obtain
\begin{align}
    \langle M_l^{x}(\tau)M_l^{x}(0)\rangle &  = \bar{C}_l \sin^{2}(2\theta) \notag \\
     + &\bar{D}_l \cos^{2}(2\theta)(\cos\Omega_l\tau+i\gamma ^{l}\sin\Omega_l \tau) \\
    \langle M_l^{y}(\tau)M_l^{y}(0) \rangle 
    &= \bar{D}_l (\cos\Omega_l\tau+i\gamma ^{l}\sin\Omega_l \tau) \\
    \langle M_l^{z}(\tau)M_l^{z}(0)\rangle &=\bar{C}_l \cos^{2}(2\theta) \notag \\ 
    + & \bar{D}_l \sin^{2}(2\theta)(\cos\Omega_l \tau+i\gamma^{l}\sin\Omega_l \tau)\
\end{align}
where 
\begin{align}
\bar{C}_l \equiv \frac{1-\gamma^{l}}{2}\, , \quad \bar{D}_l \equiv \frac{1-\gamma^{l}}{4(1+\gamma^{l})} \, .
\label{eq_mag_amp}
\end{align}

In Fig. \ref{fig_ac_mag}, we plot the above angular-time correlations for $l=4$, where we can see very similar behaviors to Fig. \ref{fig_corr_0}.
The frequency $\Omega_l$ of the oscillating term is of course identical to Eq. (\ref{eq_Omega}).
Meanwhile,  Eq. ({\ref{eq_mag_amp}}) reveals that the amplitude rapidly approaches $ \bar{D}_l \to 1/4$ for $l \gg 1$. 
In addition, it would be interesting to point out  
\begin{align}
 \sum_{\alpha} \langle (M_l^{\alpha}(0) )^2 \rangle = \bar{C}_l + 2\bar{D}_l \xrightarrow{l \to \infty  } \frac{3}{4}\, .
\end{align}
This equation implies that, for sufficiently large $l$, $M_l^\alpha$ effectively behaves as an $S=1/2$ spin, although it consists of the physical $S=1$ spins in the system part.
Thus, the angular-time correlations for $M_l^\alpha$ would work as a stable observation protocol of the entanglement spectrum associated with the edge spin degrees.
% than Eqs. (\ref{eq_aciix}), (\ref{eq_aciiy}) and (\ref{eq_aciiz}).

\section{Angular-time evolution of $S=1$ spin operators}
\label{tag_sec4}

In the previous section, we have obtained the angular-time spin correlations.
However, the angular time $\tau$ was introduced as a parameter time associated with the entanglement Hamiltonian ${\cal K}^{(l)}$, which is eventually defined in the auxiliary space rather than the physical $S=1$ spin space.
Thus, the physical interpretation of the angular-time evolution is still unclear in terms of the physical $S=1$ spins.
This point is an intrinsic difference from the original Unruh effect for a quantum field theory where the angular time is realized as the proper time of a constantly accelerating observer.
For the purpose of understanding the angular-time evolution in quantum spin systems, it is primarily important to clarify how $S^{\alpha}_{n}(\tau)$ behaves in terms of the physical $S=1$ spins.

A central issue is the property of the angular-time evolution operator ${\cal U}(\tau)$.
In order to analyze  ${\cal U}(\tau)$ in the physical spin space, it is helpful to introduce the $\tau$-dependent projection matrix,  
\begin{align}
 P(\tau) \equiv 
    &  \sum_{\nu=\pm} \langle \sigma_1 \cdots \sigma_l |\nu \rangle\!\rangle e^{-i\tau \omega_\nu} \langle\!\langle \nu |\sigma_1 \cdots\sigma_l \rangle  = \, \nonumber \\
   &\quad 
\vcenter{ \hbox{\includegraphics[scale=0.35]{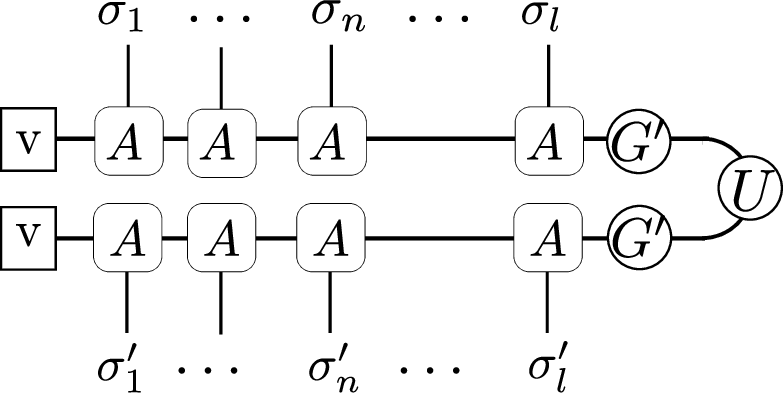}} } \; .
\end{align}
which was inserted in the core part of Eq. (\ref{eq_diagram_atc}).
Here, we remark that $P(\tau)$ can be viewed as a $\tau$-dependent generalization of the projection operator (\ref{eq_approx_trans}), which extracts the ground-state subspace from the $3^l$-dimensional physical spin space.
Note that $P(\tau)$ is reduced to Eq.(\ref{eq_approx_trans}) at $\tau=0$.

\begin{figure}
\includegraphics[scale=0.35]{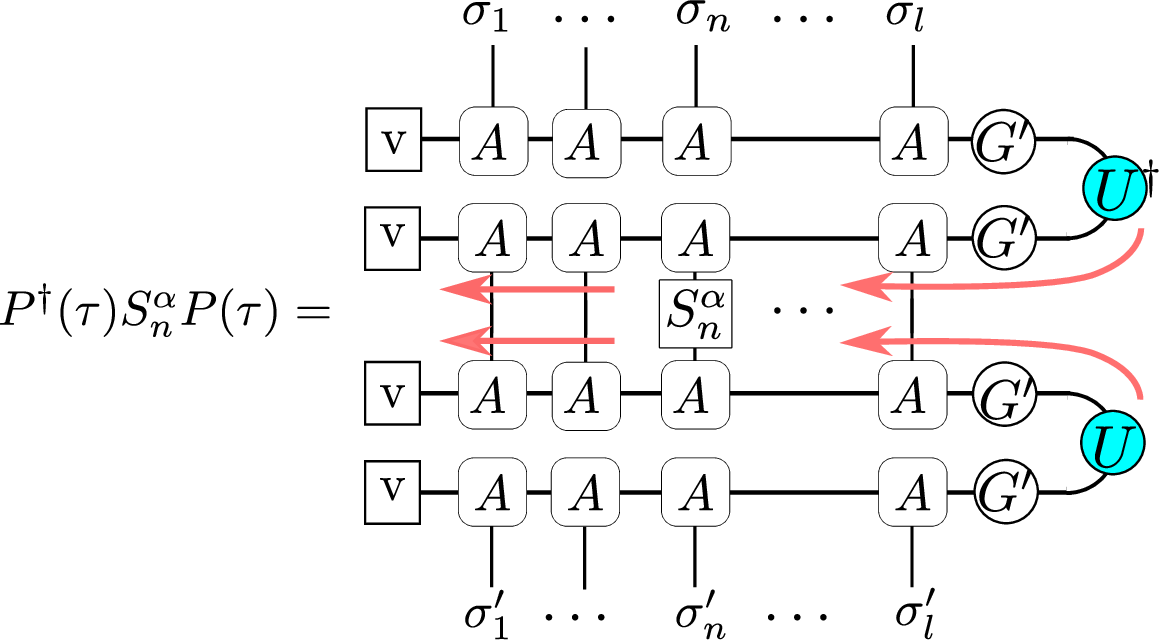}
\caption{Effective $S=1$ spin operator with the angular-time evolution. 
The arrows represent that the time evolution operator $U$ in the auxiliary space can be gauged out toward the edge sites attached with ${\rm v}$. 
}
\label{fig_atespinop}
\end{figure}

In order to figure out the physical meaning of the angular-time evolved spin in the $S=1$ spin space, we then analyze the nature of
\begin{align}
 \tilde{S}^\alpha_{n}(\tau) &=  P^\dagger(\tau)  S^{\alpha}_{n} P(\tau)  \, .
\label{eq_diagram_stau}
\end{align}
Figure \ref{fig_atespinop} shows a diagrammatic representation of Eq. (\ref{eq_diagram_stau}), where the angular-time evolution operator $U$ for the auxiliary space is located at the deep inside from the left edges of the diagram. 
In order to extract the role of $U$,  we can use the gauge transformation for the MPS, which is extensively used in the entanglement analysis of SPT states;
The SPT nature of the AKLT chain nicely enables us to move $U$ toward the left edges passing through $S^\alpha_n$ at $n$th site,

More precisely, the gauge transformation at the level of a local matrix $A^\sigma$ is illustrated as
\begin{align} 
\vcenter{ \hbox{\includegraphics[scale=0.35]{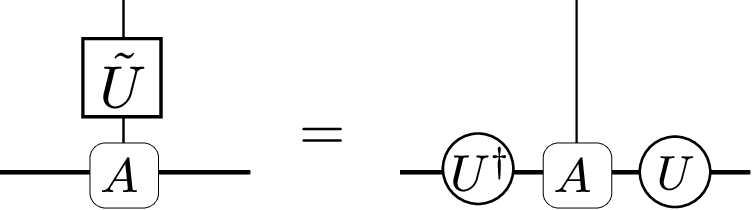}} } \; ,
\label{eq_diagram_gaugeU}
\end{align}
where a unitary transformation $U$ for the auxiliary space on the right-hand side can be converted to a transformation for the physical $\sigma$-spin space on the left-hand side.
Here, $U$ is a $2\times 2$ unitary matrix and $\tilde{U}$ denotes a $3\times 3$ unitary matrix.
For the present case, moreover, we have assumed $\theta=0$, so that $G'$ is a diagonal matrix and $[G', U] = 0$.
Using the gauge transformation (\ref{eq_diagram_gaugeU}), thus, we can actually gauged out the time-evolution matrix $U$ on both sides of $S_n^\alpha$ toward the boundary sites,  as shown by the red arrows in Fig. \ref{fig_atespinop}. 
For $U$ shifted to the left edges, moreover, it is easily seen that the $\tau$-dependence for $\mathrm{v} U = e^{-i\tau \omega_+} \mathrm{v}$ and $\mathrm{v} U^\dagger = e^{ i\tau \omega_+} \mathrm{v}$ are canceled with each other.
Thus, the effective $S=1$ operator with the angular-time evolution can be written as
\begin{align}
\tilde{S}^{\alpha}_n(\tau) = P^\dagger(0) { \mathcal{S} }_n^\alpha(\tau) P(0) \, 
\label{eq_spinrep1}
\end{align}
where 
\begin{align}
 \mathcal{S}_n^\alpha(\tau)  \equiv \tilde{U}^\dagger(\tau) S_n^\alpha \tilde{U}(\tau) \, ,
 \label{eq_spinevo}
\end{align}
and
\begin{align}
    \tilde{U}(\tau)=\exp\left( i(\omega_+ + \omega_-)\tau \right) \exp\left( -i\Omega_l \tau S_n^{z}  \right) \, .
    \label{eq_atmeansmf} 
\end{align}
The overall phase in Eq. (\ref{eq_atmeansmf}) is canceled in Eq. (\ref{eq_spinevo}).
From these, it follows that the angular-time evolved spin is effectively described by
\begin{align}
\mathcal{S}^{\alpha}_n(\tau) = e^{i\Omega_l \tau S_n^{z} }  S_n^\alpha e^{ -i\Omega_l \tau S_n^{z}  } \, .
\label{eq_effspinevo}
\end{align} 
An important feature of Eq. (\ref{eq_effspinevo}) is that it is independent of site index $n$, which also leads us to a compact representation of the angular-time evolution for the magnetization, 
\begin{align}
\tilde{M}^{\alpha}_l(\tau) \equiv  P^\dagger(\tau) M_l^\alpha P(\tau) =  P^\dagger(0) { \mathcal{M} }_l^\alpha(\tau) P(0) \, 
\label{eq_Mrep1}
\end{align}
with 
\begin{align}
\mathcal{M}^{\alpha}_l(\tau) = e^{i\Omega_l \tau  M^{z}_l  }  M_l^\alpha e^{ -i\Omega_l \tau M^{z}_l  } \, .
\label{eq_effMevo}
\end{align} 
For the XXZ chain, an angular-time evolution of the physical spin generates nonlocal terms at a finite $\tau$.
The simple form of Eqs. (\ref{eq_effspinevo}) and (\ref{eq_effMevo}) reflects the fact that the entanglement in the VBS state originates from the SPT nature of the AKLT chain.

From the viewpoint of quantum spin physics,  Eq.  (\ref{eq_effspinevo}) interestingly provides a physical interpretation of the angular-time evolution on the $S=1$ spin operator.
Eq. (\ref{eq_effspinevo}) can be equivalent to the real-time evolution of a $S=1$ spin generated by the effective uniform magnetic field in the $z$ direction
\begin{align}
\mathcal{H}_\mathrm{Z} = - h_\mathrm{eff} \sum_i S_i^z   \, ,
\label{eq_effh}
\end{align}
with $h_\mathrm{eff} \equiv \Omega_l $.
Indeed, we have confirmed that Eq. (\ref{eq_spinrep1}) repoduces Eqs. (\ref{eq_aciix}), (\ref{eq_aciiy}), and (\ref{eq_aciiz}).
Moreover, an important point in (\ref{eq_spinrep1}) and (\ref{eq_Mrep1}) is that $P(0)$ does not affect the frequency $\Omega_l$.
These suggest that although $\tau$ was originally introduced as a parameter time governed by $\mathcal{K}^{(l)}$, the angular-time evolution can be realized as real-time dynamics generated by the uniform magnetic field of Eq. (\ref{eq_effh}).

In contrast, we next discuss how the uniform magnetic field reflects on the angular-time evolution.
Since $[{\cal H}_\mathrm{AKLT}, {\cal H}_\mathrm{Z}]=0$, a weak uniform magnetic field does not disturb the ground-state wavefunction.\cite{weakfield}
Then, $\mathcal{H}_Z$ yields spatially uniform dynamical phases to the S=1 spins, which are converted into the auxiliary $S=1/2$ spin space.
% in a spatially nonuniform way.
The $l$-dependent frequency $\Omega_l$ explains how such dynamical phases emerge at the entangle point at the $l$-th site from the left edge.
Here, we should also remark that if the magnetic field $h_\mathrm{eff}\to h$ in Eq. ({\ref{eq_effh}) is continuously changed, the autocorrelation function of Eq. (\ref{eq_effspinevo}) is given by Eqs. (\ref{eq_aciix}), (\ref{eq_aciiy}), and (\ref{eq_aciiz}) with replacing $\Omega_l\to \Omega(=h)$.
This suggests that the angular-time spin correlations do not exhibit any particular behavior at $h=\Omega_l$, which makes it hard to identify $\Omega_l$ with such a magnetic resonance experiment.

\section{Summary and Discussions}

In this paper, we have introduced the angular-time evolution for quantum spin chains through the entanglement Hamiltonian of the biparititioned ground-state wavefunction, in analogy with the Rindler-time evolution in the Unruh effect for quantum field theories.
In particular, we have investigated the angular-time evolution for the ground state of the half-infinite Affleck-Kennedy-Lieb-Tasaki (AKLT) chain with the open boundary condition, where the exact eigenvectors of the reduced density matrix can be explicitly constructed with the matrix-product state (MPS).
We have analytically calculated the angular-time spin correlations and then found that they exhibited the oscillating behavior with the frequency $\Omega_l$ defined by Eq. (\ref{eq_Omega}),  reflecting the entanglement spectrum for the system part of length $l$.
Using the gauge transformation for the MPS, moreover, we have clarified that the angular-time evolution can be realized as edge-state dynamics induced by the uniform magnetic field of Eq. (\ref{eq_effh}) for $S=1$ spins, although the angular time $\tau$ was at first introduced as a dimensionless parameter due to $\mathcal{K}^{(l)}$ in the purely theoretical context.

The nontrivial entanglement in the ground state of the  AKLT chain originates from the SPT order due to the $\mathbb{Z}_2\times\mathbb{Z}_2$ symmetry.
Thus, how the angular-time evolution is converted into that for the classical state by the Kennedy-Tasaki nonlocal transformation that disentangles the global SPT entanglement structure may be a theoretically important question.\cite{KT_prb1992, Oshikawa1992, Okunishi2011,Pollmann2012,Okunishi2014}
As discussed in Sec. \ref{tag_sec4}, moreover, the nature of the SPT state enables us to relate the angular-time evolution with the real-time dynamics of the $S=1$ spins in the uniform magnetic field.
Here, we would like to remind that an electron-spin-resonance experiment interestingly captured the resonances of edge spin modes associated with Haldane states of finite lengths for Y$_2$BaNi$_{0.96}$Mg$_{0.04}$O$_5$, which is described as an assembly of the $S=1$ Heisenberg chains of various chain-lengths with small anisotropies.\cite{YoshidaESR, Batista1998,Batista1999}
It is an intriguing problem to discuss the angular-time evolution for the $S=1$ Heisenberg chain with small anisotropies.

In addition, recently, tomography protocols of the entanglement Hamiltonian that contains spatially nonuniform interactions were proposed  for trapped-ion simulators.\cite{Kokail2021} 
A couple of measurement protocols of the spectrum associated with the SPT entanglement were suggested for ultra-cold atoms or quantum circuit models. \cite{ES_Measurement, Choo2018}
Edge states of 2D AKLT models may also provide a research stage for the SPT entanglement.\cite{Lou2011,Liu2022}
In such systems, ``real-time" dynamics of entanglement Hamiltonians would be experimentally observable in principle.
Thus, it is also an interesting problem from both theoretical and experimental points of view to investigate how the angular-time evolution can be realized in such experimentally relevant situations.

\acknowledgments 

This work is supported by Grants-in-Aid for Transformative Research Area "The Natural Laws of Extreme Universe---A New Paradigm for Spacetime and Matter from Quantum Information" (Grant No. JP21H05182 and No. JP21H05191) from JSPS of Japan.

\appendix
% \section{CONSTRACTION OF MPS REPRESENTATION OF VBS STATE}\label{app:mpsrep}

\section{Matrix elements for the $S=1$ spin matrices}\label{app_A}

%\subsection{matrix elements}

We calculate  matrix elements of $S_n^\alpha$ with the $ |q \rangle\!\rangle$ bases in the auxiliary space with the MPS formalism, which are diagrammatically illustrated as
\begin{align}
\langle \!\langle  q |S_n^\alpha| q' \rangle \! \rangle  = \; 
\vcenter{ \hbox{ \includegraphics[scale=0.40]{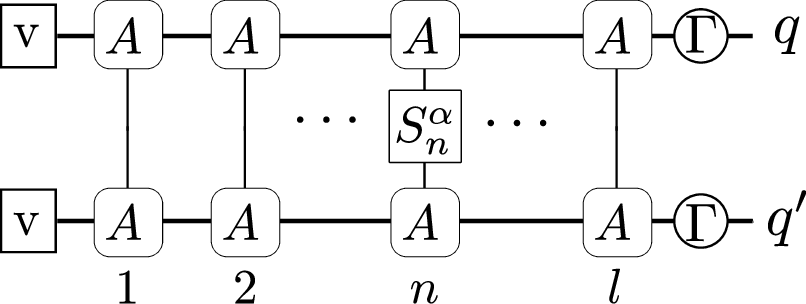} } } 
\end{align}
with $q, \, q' = \uparrow\, , \downarrow$.
The resuls are summarized as the following $2\times 2$ matrices:
% we need $\braket{v_{j}|S^{a}_{i}|v_{k}}$ and $\braket{q=\uparrow,\downarrow|S^{a}_{i}|q=\uparrow,\downarrow}$.
\begin{align}
S^{x}_{n}  \, &: \; 
\begin{bmatrix}
-\gamma^{n} \sin(2\theta) & - \gamma^{l-n +1 } \\
-\gamma^{l-n+1}& \gamma^{n} \sin(2\theta)
\end{bmatrix}
\, , \\
S^{y}_{n} \; &: \;
\begin{bmatrix}
0   & -i\gamma^{l-n+1}  \\
i \gamma^{l-n+1}  & 0
\end{bmatrix}\, , \\ 
%\end{align}
%\begin{align}
S^{z}_{n} \, &: \; 
\begin{bmatrix}
-\gamma^{l-n+1} -\gamma^n \cos2\theta & 0 \\
0 & \gamma^{l-n+1}-\gamma^n \cos2\theta 
\end{bmatrix}
\, ,
\end{align}
where the row and column indices respectively correspond to $q$ and $q'$.

\section{The angular-time spin correlations for separated spins} 
\label{sec_appdx2}

We calculate $\langle S^{\alpha}_{n}(\tau)S^{\alpha}_{m}(0)\rangle $ for $1 \le n, m \le l$ using the MPS formulation, using the matrix elements in Appendix A.
The results are summarized as 
\begin{align}
\langle S^{x}_{n}(\tau)&S^{x}_{m}(0)\rangle  =  \tilde{C}_{l,n,m} \sin ^2(2 \theta ) \notag\\
 &+ \tilde{D}_{l,n,m} \cos^2(2\theta)(\cos\Omega_l\tau+i\gamma^{l}\sin\Omega_l \tau)\, ,\\
\langle S^{y}_{n}(\tau)&S^{y}_{m}(0) \rangle =\tilde{D}_{l,n,m}(\cos\Omega_l\tau+i\gamma^{l}\sin\Omega_l\tau) \, , \\
\langle S^{z}_{n}(\tau)&S^{z}_{m}(0)\rangle  = \tilde{C}_{l,n,m} \cos ^2(2 \theta ) \rangle\notag\\
 &+ \tilde{D}_{l,n,m} \sin^2(2\theta)(\cos\Omega_l\tau+i\gamma^{l}\sin\Omega_l\tau) \, 
\end{align}
where
\begin{align}
\tilde{C}_{l,n,m} & = 
 \frac{2 \left(\gamma^{n}-\gamma^{l-n+1}\right)\left(\gamma^{m}-\gamma^{l-n+1}\right)}{1+\gamma^l} \notag \\
&+\frac{2 \left(\gamma^{ n}+\gamma^{l-n+1}\right)\left(\gamma^{ m}+\gamma^{l-m+1}\right)}{1-\gamma^l} \label{eq_appb_C}\\
\tilde{D}_{l,n,m}& = \frac{4\gamma^{2l-n-m+2}}{1-\gamma^{2l}}\, \label{eq_appb_D},
\end{align}
Note that $C_{l,n}$ and $D_{l,n}$ in Eqs. (\ref{eq_C_ln}) and  (\ref{eq_D_ln}) are respectively reproduced if $n=m$ in Eqs. (\ref{eq_appb_C}) and (\ref{eq_appb_D}).

\bibliography{nakajima}

\begin{thebibliography}{49}
\bibitem [{wea()}]{weakfield}%
  \BibitemOpen
  \href@noop {} {}\bibinfo {note} {The AKLT chain has a large excitation gap of
  about $0.71J$ with $J$ being the exchange coupling. Thus, a weak uniform
  magnetic field does not affect the ground state of the AKLT
  chain.}\BibitemShut {Stop}%
\end{thebibliography}

\end{document}